\documentclass[preprint]{aastex} 
\usepackage{natbib}

\shorttitle{TAOS: Stellar Variability II }
\shortauthors{Mondal et al.}

\begin{document}
 
\title{The TAOS Project Stellar Variability II. Detection of 15 Variable Stars}

\author{ 
S.~Mondal\altaffilmark{1,2}, 
C.C.~Lin\altaffilmark{1}, 
W. P.~Chen\altaffilmark{1},
Z.-W.~Zhang\altaffilmark{1},  
C.~Alcock\altaffilmark{3},
T.~Axelrod\altaffilmark{4},
F.~B.~Bianco\altaffilmark{3,5,6,7},
Y.-I.~Byun\altaffilmark{8},
N.~K.~Coehlo\altaffilmark{9},
K.~H.~Cook\altaffilmark{10},
R.~Dave\altaffilmark{11},
D.-W.~Kim\altaffilmark{8},
S.-K.~King\altaffilmark{12},
T.~Lee\altaffilmark{12},
M.~J.~Lehner\altaffilmark{12,5,3},
H.-C.~Lin\altaffilmark{1},
S.~L.~Marshall\altaffilmark{13,10},
P.~Protopapas\altaffilmark{11,3},
J.~A.~Rice\altaffilmark{9},
M.~E.~Schwamb\altaffilmark{14},
J.-H.~Wang\altaffilmark{1,12},
S.-Y.~Wang\altaffilmark{12} and
C.-Y.~Wen\altaffilmark{12}
}
\altaffiltext{1}{Institute of Astronomy, National Central University,
 300 Jhongda Rd, Jhongli 32054, Taiwan}
\email{soumen@aries.res.in}
\altaffiltext{2}{Aryabhatta Research Institute of Observational
  Sciences, Manora Peak, Nainital-263129, India}
\altaffiltext{3}{Harvard-Smithsonian Center for Astrophysics, 60 Garden Street,
 Cambridge, MA 02138}
\altaffiltext{4}{Steward Observatory, 933 North Cherry Avenue, Room N204
 Tucson AZ 85721}
\altaffiltext{5}{Department of Physics and Astronomy, University of
 Pennsylvania, 209 South 33rd Street, Philadelphia, PA 19104}
\altaffiltext{6}{Department of Physics, University of California Santa Barbara, Mail
  Code 9530, Santa Barbara CA 93106-9530}
\altaffiltext{7}{Las Cumbres Observatory Global Telescope Network, Inc. 6740 Cortona Dr. Suite
  102, Santa Barbara, CA 93117}
\altaffiltext{8}{Department of Astronomy, Yonsei University, 134 Shinchon,
  Seoul 120-749, Korea}
 \altaffiltext{9}{Department of Statistics, University of California Berkeley,
 367 Evans Hall, Berkeley, CA 94720}
\altaffiltext{10}{Institute for Geophysics and Planetary Physics, Lawrence
 Livermore National Laboratory, Livermore, CA 94550}
 \altaffiltext{11}{Initiative in Innovative Computing at Harvard,
 120 Oxford St., Cambridge MA 02138}
\altaffiltext{12}{Institute of Astronomy and Astrophysics, Academia Sinica.
 P.O. Box 23-141, Taipei 106, Taiwan}
\altaffiltext{13}{Kavli Institute for Particle Astrophysics and Cosmology,
 2575 Sand Hill Road, MS 29, Menlo Park, CA 94025}
\altaffiltext{14}{Division of Geological and Planetary Sciences,
 California Institute of Technology, 1201 E. California Blvd., Pasadena, CA
 91125}

\begin{abstract}
The Taiwanese-American Occultation Survey (TAOS) project has collected
more than a billion photometric measurements since 2005 January. These
sky survey data---covering timescales from a fraction of a second to
a few hundred days---are a useful source to study stellar
variability.  A total of 167 star fields, mostly along the ecliptic
plane, have been selected for photometric monitoring with the TAOS
telescopes.  This paper presents our initial analysis of a search for
periodic variable stars from the time-series TAOS data on one
particular TAOS field, No.\,151 (RA = 17$^{\rm h}$30$^{\rm m}$6$\fs$67,
Dec = 27\degr17\arcmin 30\arcsec, J2000), which had been observed over
47 epochs in 2005. A total of 81 candidate variables are identified in
the 3 square degree field, with magnitudes in the range $8 < R < 16$.
On the basis of the periodicity and shape of the lightcurves, 29   
variables, 15 of which were previously unknown, are classified as RR
Lyrae, Cepheid, $\delta$ Scuti, SX Phonencis, semi-regular and
eclipsing binaries.
\end{abstract}


\section{Introduction} %1 

The Taiwanese-American Occultation Survey (TAOS) project aims to
search for stellar occultation by small ($\sim$1~km diameter)
\emph{Kuiper Belt Objects} (KBOs). The KBO population consists of
remnant planetesimals in our Solar System, which typically have low to
intermediate (below 30\degr) inclination orbits and heliocentric
distances between 30 and 50~AU \citep{edg49,kui51,mor03}. The size
distribution of large KBOs shows a broken power law with the break
occurring $\sim$30--100~km that indicates a relative deficiency of small
KBOs. Such a broken power law is believed to be the consequence of
competing processes of agglomeration to form progressively larger
bodies versus collisional destruction. The size distribution thus
provides critical information of the dynamical history of the Solar
System.  The stellar occultation technique, namely the dimming of a
background star by a passing KBO, is the only technique capable of
detecting cometary-sized bodies, which are too faint for direct
imaging even with the largest telescopes \citep{alc03,zha08}.  So far,
TAOS has collected several billion stellar photometric measurements,
and no occultation events have been detected, indicating a significant
depletion of small KBOs \citep{zha08,bia09}.

Several projects have discovered numerous variable stars as
byproducts, for instance the MACHO \citep{alc95, alc98}, EROS
\citep{bea95,der02}, OGLE \citep{cie03, wra04}, and ROTSE-I
\citep{aker00,kin06,hof09}.  Such data have enriched our knowledge of
stellar variability in the Galactic fields and the Magellanic Clouds,
which not only improves the number statistics, but also has helped to
shed light on the detailed mechanisms of stellar variability.
Knowledge of the variability has been so far still relatively poor for
even the bright stars.  Recent large-area sky survey projects, however, 
have started to turn up large numbers of variable stars.  These projects include 
the All Sky Automated Survey \citep[ASAS]{poj05}, the observations by the 
Hungarian Automated Telescope \citep[HAT]{bak01}, the Northern Sky Variability Survey 
\citep[NSVS]{woz04}, and ROTSE-I.  Variable stars, notably Cepheids, 
RR Lyrae-type, $\delta$ Scuti-type, SX Phonenicis-type, semi-regular variables, 
and eclipsing binaries are shown to be ubiquitous in Galactic fields and in clusters.  
The next-generation projects like the cyclic all-sky survey by the Panoramic Sky 
Survey And Rapid Response System (Pan-STARRS) no doubt will provide a much 
complete variable star census and characterization to enhance vastly 
our understanding of the cosmos in the time domain.  

While the main goal of the TAOS project is to conduct a KBO census by
detecting stellar occultations, the plethora of time-series stellar
photometry renders the opportunity to identify and characterize
variable stars spanning a wide range of timescales, from less than a
second to a few years.  The first paper of the series of the TAOS stellar  
variability studies deals with detection of low-amplitude $\delta$ 
Scuti stars \citep{kim10}.  The current paper, the second in the series, presents 
the effort to identify variable stars in a targeted star field.    

\section{Observations and Data Reduction} %2

The TAOS telescope system consists of an array of four 50~cm, fast
optics (f/1.9), wide-field robotic telescopes, sited at Lulin
Observatory (longitude 120$^{\circ}$ 50\arcmin ~28\arcsec E; latitude
23$^{\circ}$ 30\arcmin ~N, elevation 2850 meters) in central Taiwan.
Each telescope is equipped with a 2048 $\times$ 2048 SI-800 CCD
camera, with a pixel scale of 2.9\arcsec, yielding about a 3 square
degree field of view on the sky.  The TAOS system uses a broad
custom-made filter which, together with the sensitivity of the CCD,
has a response function close to that of a standard broad R-band
filter.

All TAOS telescopes observe the same star field simultaneously so as
to eliminate false detection of occultation events by KBOs.  Each
observing session begins with regular imaging (``\emph{stare mode}'')
of the star field, followed by a special CCD readout operation
(``\emph{zipper mode}'').  In the zipper mode, the camera continues to
read out a block of pixels at a time while the shutter remains open.
A stellar occultation by a km-sized KBO is expected to last for only a
fraction of a second, and it is this pause-and-shift charge transfer
operation that allows 5~Hz photometric sampling to detect such an
event.  The zipper-mode data are most suitable to study truly fast
varying phenomena such as stellar flaring, but they were not used in
the results reported in this paper so will not be discussed further.
Technical details of the TAOS operation can be found in \citet{leh09}.

The primary purpose of the stare-mode observations is to provide
guidance of the pointing of the star field, particularly for
photometric processing of the zipper-mode images, but the stare-mode
data can be used also for stellar variability studies.  A set of
stare-mode observations consists of 9 telescope pointings, each with 3
frames of images, dithered around the center of target field.  The 
frames covering the central position were used in the analysis
reported here.

There are a total of 167 TAOS star fields, mostly along the ecliptic
plane.  These fields have been selected to have few exceedingly bright
($R< 7$) stars, and to have a sufficient number of stars to maximize
occultation probability, yet not too crowded to hamper accurate
stellar photometry.  The number of stars brighter than about $R\sim
16$~ranges from a few hundreds to several thousands in each of our
target fields.

This paper presents the variable stars found in a particular field,
No.\,151, which has the central coordinates RA = 17$^{\rm h}$30$^{\rm
  m}$6$\fs$67, Dec = 27\degr17\arcmin30\arcsec (J2000).
After excluding data taken under inferior sky conditions, the data
presented here include 93 good photometric measurements taken at 47
epochs from 2005 April 11 to 2005 August 02.  Each photometric
measurement came from a stare-mode image with a 4~s exposure.

Photometry was performed using the \emph{SExtractor} package
\citep{sex96} with a $3\sigma$ source detection limit. For each
detected source, the output provides the x-y position, instrumental
magnitude, magnitude error, FWHM, etc.  Astrometry was done using
\emph{imwcs} task of \emph{WCSTools}\footnote[1]{Package available at
  http://tdc-www.harvard.edu/software/wcstools/} \citep{min99} with
the USNO-B1.0 catalog \citep{mon03}.  Then the CCD x-y positional
output from the {\it SExtractor} was converted to sky coordinates (RA
and DEC) for individual images using the \emph{xy2sky} task of
\emph{WCSTools}.

The stellar position was matched with the USNO-B1.0 and 2MASS
\citep{cut03} catalogs. The USNO-B1.0 catalog was derived from images
of digitization of sky-survey photographic plates, and gives
the position, proper motions, photographic magnitude in each of the
five passbands (B1, B2, R1, R2, I), and star/galaxy estimators for
some 1,042,618,261 objects.  The 2MASS Point Source Catalog
essentially covers the whole sky in three near-infrared bands J, H and
K$_s$, down to a limiting magnitude of $J\sim15.8$~mag with a
signal-to-noise ratio of 10.  Optical magnitudes, USNO unique
identification numbers (USNO ID), and 2MASS magnitudes of the detected
sources in our images were obtained by matching the position to the
USNO-B1.0 and 2MASS catalogs.  A matching radius of 10$\arcsec$ was
used, which gives unambiguous identifications in all but a few cases.
The catalog for each image provides the unique USNO\_id, TAOS
instrumental magnitude, optical magnitude (B2, R2), 2MASS\_id, and
infrared (J, H, K$_s$) magnitudes.  

We then created the lightcurve for each star, containing the modified 
Julian date (MJD), calibrated TAOS magnitude, and error in magnitude.  
Photometric calibration of the TAOS instrumental magnitude will be 
discussed in the next section.  Only data with good photometric quality, 
judged on the basis of the number of detected sources, were used in the 
analysis.  Best images are those with more than 3000 detected sources. 
In the results reported here, we only
considered images having more than 2500 detected sources.  For
variability analysis, only sources with more than 80 photometric
measurements were considered.  Finally we had the lightcurves of 2915
sources, mostly with 93 photometric measurements.

\subsection {Photometric Calibration} %2.1

\begin{figure*}
% \plotone{figure1.ps}
 \center
 \includegraphics[angle=-90, width=\textwidth]{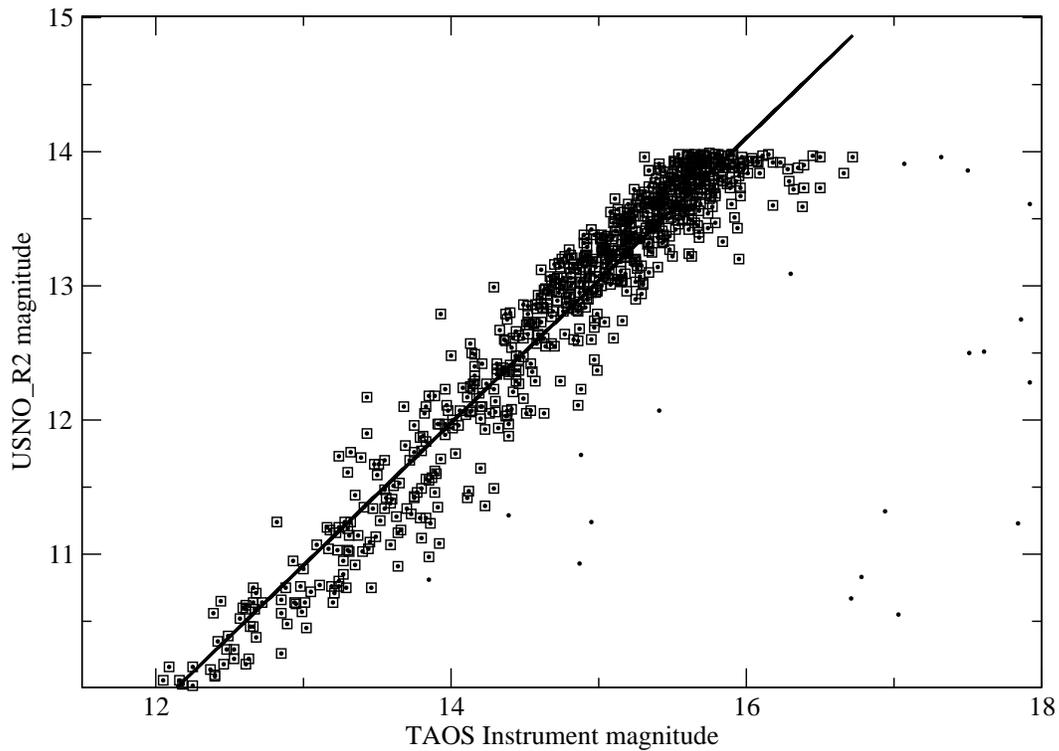}
\caption{The TAOS instrument magnitude versus the R2 magnitude in the
  USNO-B1.0 catalog.  The dots shows all the TAOS measurements, whereas the 
  squares mark those stars with a corresponding R2 magnitude between 10 and 14 
  used in the linear fitting, shown as the solid line.  Outliers, i.e., TAOS 
  detections with mismatched USNO magnitudes, are caused by the photometric scattering 
  of the USNO-B1.0 photometry or bad pixels in the TAOS images. }
\label{mag:cal}
\end{figure*}

TAOS images are obtained with a filter close (but not identical) to the
standard R optical band.  We used the R2 magnitude in the USNO-B1.0
catalog to calibrate our TAOS instrumental magnitude with a linear
fit, under the assumption that most stars are not variable.
Despite the large photometric scattering intrinsic to the USNO-B1.0
catalog (derived from photographic plates), the calibration gives a
consistent rescaling of the TAOS instrumental magnitude for each star
so as to remove run-to-run variable sky transparency, atmospheric extinction
due to different air-masses, and telescope system variations.  One
such calibration curve for a particular image is shown in
Figure~\ref{mag:cal}.  

\subsection{Periodicity Analysis} %2.2

We used the \emph{Lomb-Scargle} (LS) periodogram \citep{lom76,sca82}
to determine the most likely period of a variable star. The LS method
computes the Fourier power over an ensemble of frequencies, and finds
significant periodicities even for unevenly sampled data.  We used the
algorithm taken from the Starlink\footnote[2]{http://www.starlink.uk}
software database available publicly, and verified the periods further
with the software
\emph{Period04}\footnote[3]{http://www.univie.ac.at/tops/Period04}
\citep{len05} for stars displaying obvious periodic variation.
\emph{Period04} also provides the semi-amplitude of the variability in
a lightcurve.  For any star showing a possibly spurious period, we
carefully checked the phased lightcurve for that particular period.

\section{Results} %3

\subsection{Candidate Variables} %3.1 

\begin{figure*} 
 \plotone{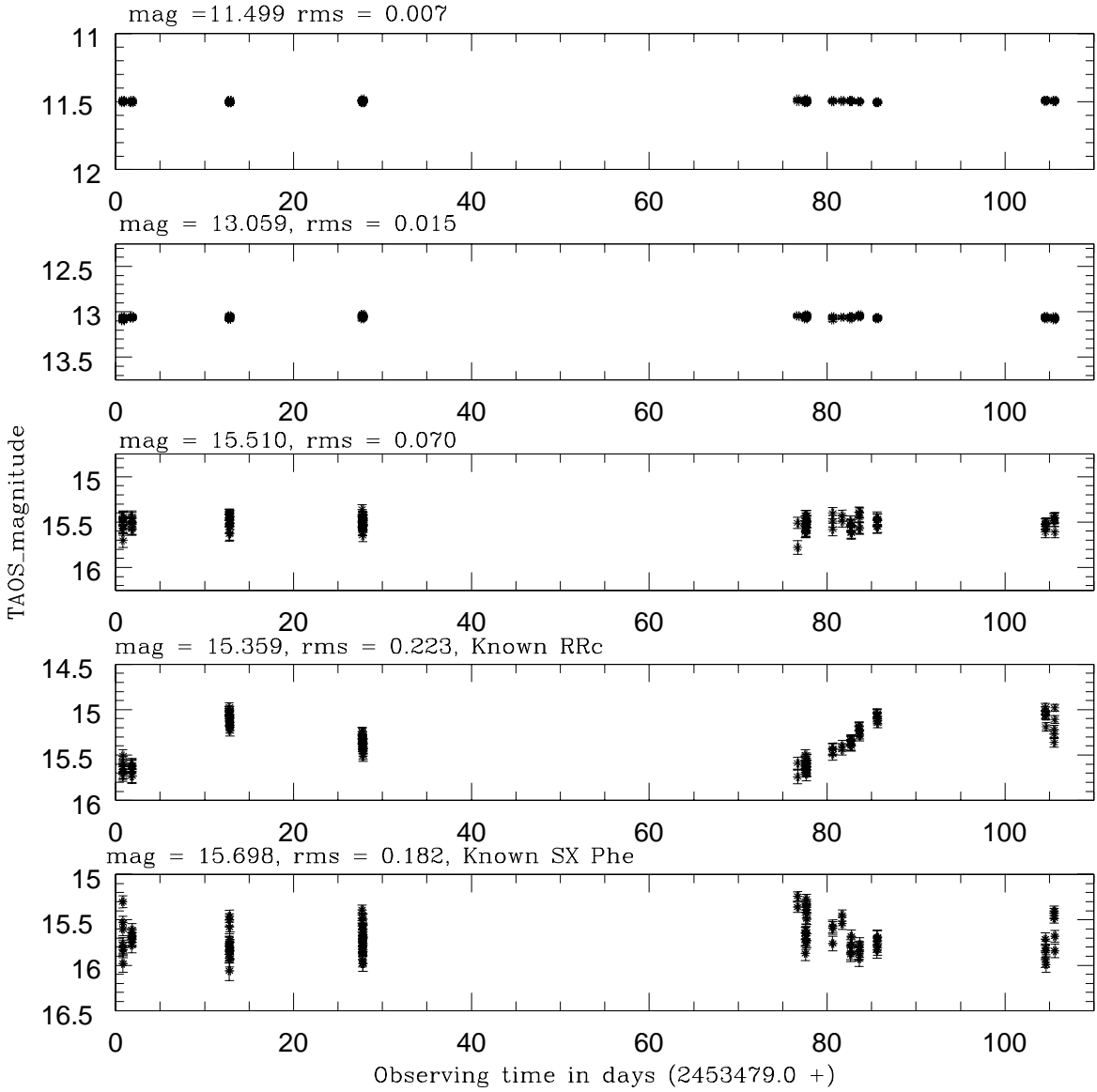} 
 \caption{Example lightcurves of a few apparently nonvariable stars 
  (top 3 panels) and known variable stars.}
 \label{nonvar}
\end{figure*}

\begin{figure*}
\plotone{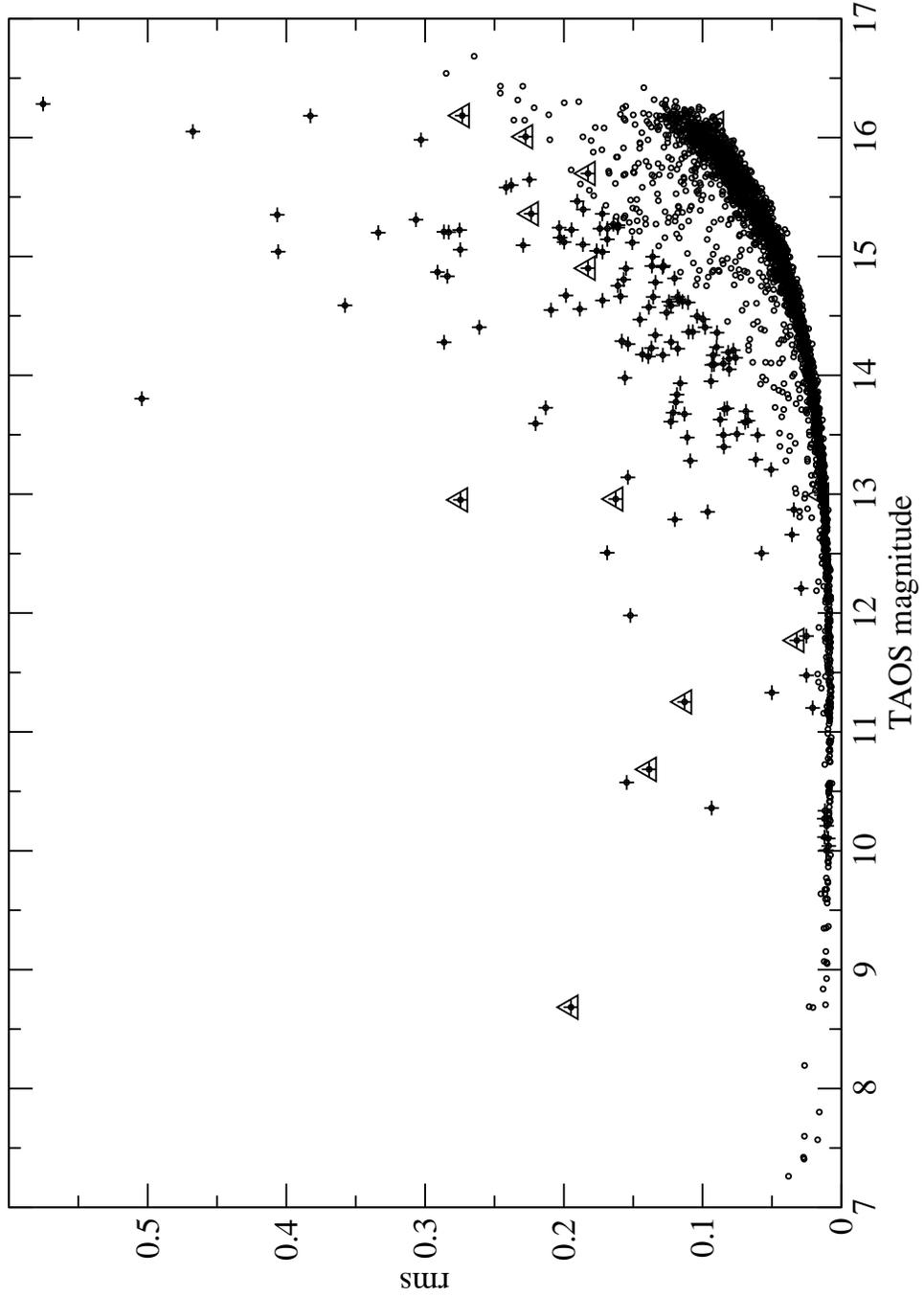}
\caption{The root-mean square of the magnitudes of some 2900 stars in
  the 2.9 deg$^{2}$ field of the TAOS Field 151 (RA = 17$^{\rm
    h}$30$^{\rm m}$6$\fs$67, Dec = 27\degr17\arcmin30\arcsec).
  Each dot represents the RMS of the lightcurve of each of the 2900
  stars in the field.  The triangles mark known variables.  The plus symbols 
  tag the sources having more than 3$\sigma$ variability in the lightcurves, 
  hence are variable candidates.}
\label{rms}
\end{figure*}

Because a large number of stars have been observed, the random errors
of the differential magnitudes are well determined.  These amount to
$\sim$0.02~mag at TAOS magnitude $\la$14 but increases to
$\sim$0.1~mag at $\sim$16~mag. To illustrate this, Figure~\ref{nonvar}
shows the lightcurves of a few nonvariable stars (per our analysis) as
well as known variable stars.  Figure~\ref{rms} shows the variations
of the lightcurves of 2900 stars in the selected field.  Each point
represents the root-mean-square (RMS) of 80--93 measurements of a
particular star over 105 days.  One sees that most stars behave
``normally'', i.e., the signal to noise decreases for fainter stars,
as expected.  The increase of RMS at the bright end ($<8$~mag) is due
to saturation.  An outlier, that is, a star with a large RMS value for
its magnitude, is then considered a likely variable.

A total of 143 variable star candidates were identified on the basis
of 3$\sigma$ above the average RMS in a magnitude bin.  Visual
inspection of the lightcurves indicated that 62 stars show large RMS
values because of flux drops of only a few data points, e.g., as the
result of bad pixels or cosmic rays.  These were excluded from the
variable list.  At the end we had the final count of 81 candidate
variables.

\subsection{Previously Known Variables} %3.2

We have searched the \emph{International Variable Star Index} 
(VSX)\footnote[4]{http://www.aavso.org/vsx/} of the American Association of Variable 
Star Observers (AAVSO) for known variables in our field.   The VSX database is 
populated with the \emph{General Catalogue of Variable Stars} (GCVS) \citep{kho98}, 
the New Catalog of Suspected Variable Stars \citep{kuk82}, and the 
published data from sky surveys, e.g., the Northern Sky Variability 
Survey (NSVS) \citep{woz04}, the All Sky Automated Survey (ASAS) \citep{poj05}, 
and the Optical Gravitational Lensing Experiment-phase 2 (OGLE-II) \citep{woz02}.  

A total of 19 variables from the VSX database are found in our field, and
we recovered 14 of them, listed in Table~\ref{known_var}.  Missing
objects in our list of known variables include one star that is brighter than 
our saturation limit so was not included in our analysis.
Two stars, namely ASAS J172907+2749.4 \citep{puj02} and
ROTSE1 J172907.35+274928.6 \citep{aker00}, with a 3.6\arcsec
coordinate difference, are classified as an RRAB with similar periods.
They should be the same star, and we recovered the object as USNO-B1 ID
1178.0358793 with a similar period.
Another two entries in
the VSX, namely ROTSE1\,J173203.69+272225.1 from ROTSE-I data
\citep{aker00} and ASAS\,J17320+2722.4 from HAT data \citep{puj02}, also
with a 3.6\arcsec difference in the published coordinates, have only one
counterpart in our data as USNO-B1 ID 1173.0334705.  They are detected
as one star by the TAOS telescope system, which has a large 3\arcsec\ 
pixel scale.  We suspect the 3.6\arcsec coordinate difference to be a 
systematic offset in the ASAS catalog, and these two entries
actually refer to the same star.  This star is classified
as a long-period variable (LPV) by ROTSE-I without period determination.
We estimated the period to be about 104 days.
       
Of the three stars, V0486\,Her, ROTSE1 J172638.42+265616.5 and ASAS
J172638+2656.3, the first two having the same coordinates, period  
(0.8059/0.8056 days) and classification (RRAB), but as separate entries
in the VSX database.  The third star, again 3.6\arcsec away, should also be
the same star, though ASAS gave a similar brightness but classified it as a 
CW-FU with a different period (4.203 days).  We recovered one variable as 
USNO ID 1169.0316280 in the position, with a period of 0.8061 days.  We hence caution 
on the possible multiple entries of the same variable star in the VSX database.  

\begin{figure*}
\plotone{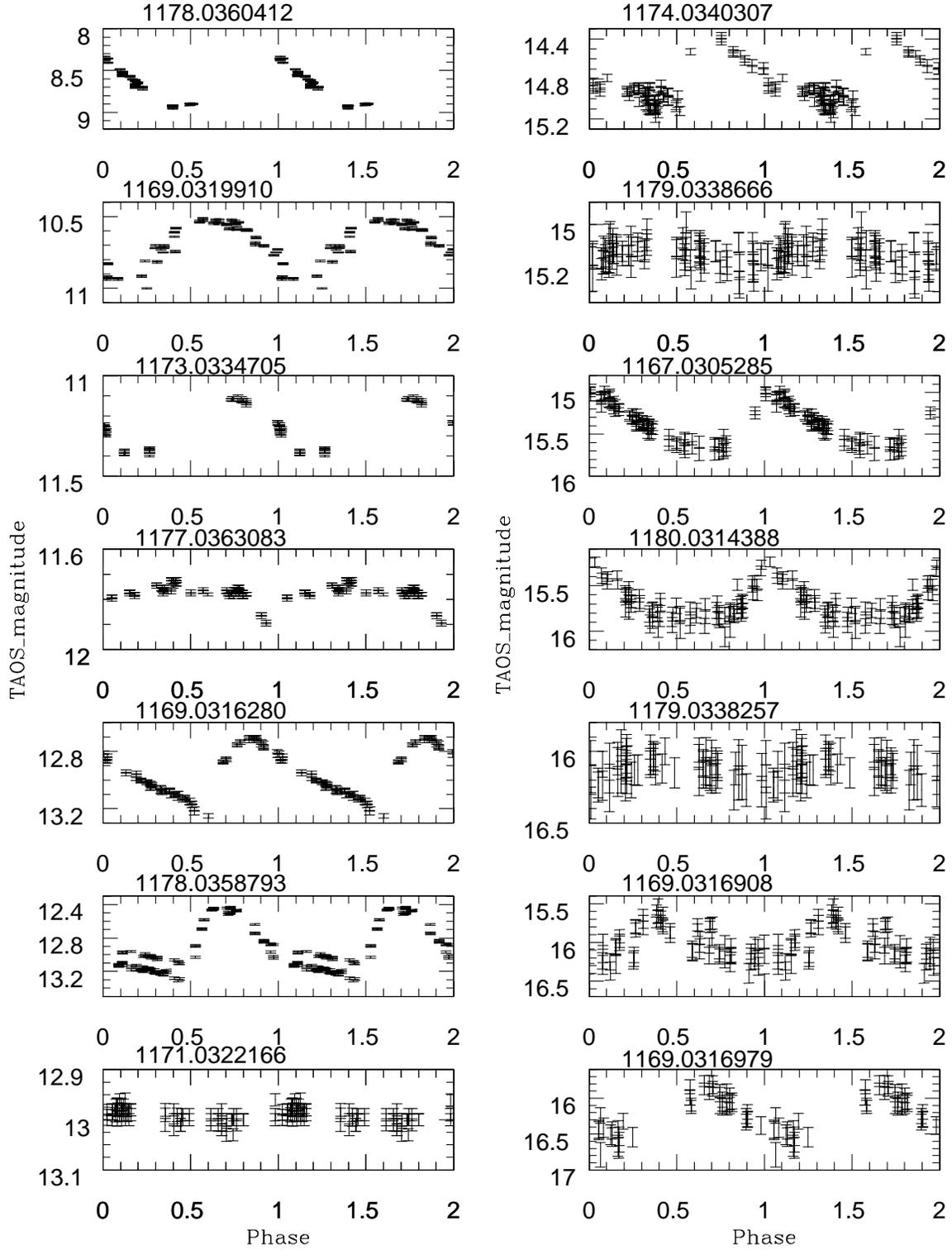}
\caption{Phased lightcurves of known variables in Table~\ref{known_var}.}
\label{fig:known_var}
\end{figure*}

The VSX database provides information on the variable type, period, and magnitude range.  
Table~\ref{known_var} lists the period and semi-amplitude of known variables derived 
from our new data, together with VSX information.  The first two columns are the star 
name and the USNO identification number (ID).  The third and fourth columns provide 
the visual magnitude range in the VSX database and our TAOS magnitude range.  
Columns 5 to 7 give, respectively, the number of observed frames, RMS in TAOS 
magnitude and the variable classification from the VSX database (such as the
pulsating types of LB and LPV, RRAB, and RRC, and the eclipsing binary types 
of EW end EA, and the $\delta$ Scuti type of SX Phe.)  Column 8 provides the 
period taken from the VSX catalog, while the ninth and tenth columns give
the periods we derived using the LS method and {\it Period04}, described in Section~2.2.
Column 11 provides the semi-amplitude determined by {\it Period04}.
The star NT\,Her is a known LB variable, but without a published period.
Our data suggest a long period, $\sim 74$~days with a large uncertainty.  For the 
rest of known variables, the periods we determined are consistent, except in 
harmonics in some cases, with those listed in the VSX.  The lightcurves of 
the 14 previously known variable stars are shown in Figure~\ref{fig:known_var}.

%found 8 ??variables in our field, 7 of which matched
%with our variable candidates list. The missing one is brighter than
%our saturation limit so was not included in our analysis.  The GCVS
%provides information on the variable type, period and visual magnitude
%at maximum.  Table~\ref{known_var} lists the period and semi-amplitude
%of known variables derived from our new data, together with GCVS
%information.  The first two columns are the star name and the
%identification number (ID) in the USNO-B1 catalog.  The third and
%fourth columns provide the visual magnitude range in the GCVS catalog
%and our TAOS magnitude range. Columns 5 to 7 give, respectively, the
%number of observed frames, RMS in TAOS magnitude and the variable
%classification from the GCVS catalog (such as the pulsating types of
%LB, RRAB, and RRC, and the eclipsing binary types of EW end EA, and
%the $\delta$ Scuti type of SX Phe.)  Column 8 provides the period
%taken from the GCVS catalog, while the ninth and tenth columns give
%the periods we derived using the LS method and \emph{Period04},
%described in Section~2.2.  Column 11 provides the semi-amplitude
%determined by \emph{Period04}.  The star NT Her is a known LB
%variable, but without a published period.  Our data suggest a long
%period, $\sim 74$~days with a large uncertainty.  For the rest of
%known variables, the periods we determined are consistent (except in
%harmonics in some cases) with those listed in the GCVS. The
%lightcurves of the 7 previously known variable stars are shown in
%Figure~\ref{fig:known_var}.

%\input  table1.tex
\begin{deluxetable}{llllcllllll}
\rotate 
\tablecolumns{11} 
\tablecaption{Known Variable Stars}
\tabletypesize{\scriptsize}
\tablewidth{0pt}
\tablehead{ 
\colhead{USNO id} & \colhead{Star Name} & \colhead{Vmag} &
\colhead{Tmag} & \colhead{Frames} & \colhead{$\sigma_T$} &
\colhead{Type} & \colhead{P$_\mathrm{known}$} & \colhead{P$_{LS}$} &
\colhead{Period04} & \colhead{Semiamp} } 
\startdata
1178.0360412 & NT Her                     & 10.0-10.6       &  8.33- 8.96 & 90 & 0.192  & LB    & NA        & 74.5240  & 74.8600  & 0.309 \\
1169.0319910 & V1097 Her                  & 10.7-11.3       & 10.51-11.00 & 88 & 0.133  & EW    & 0.3608    &  0.1804  &  0.1805  & 0.179 \\
1173.0334705 & ASAS J173204+2722.4        & 11.51(0.336)    & 11.10-11.40 & 93 & 0.113  & MISC  & -     & $\sim 104$ & $\sim 104$ & 0.19  \\ 
             & ROTSE1 J173203.69+272225.1 &                 &             &    &        & LPV   &           &          &          &       \\ 
1177.0363083 & V1060 Her                  & 12.1-12.8       & 11.72-11.90 & 92 & 0.032  & EA    & 1.5768    &  0.7012  &  1.5873  & 0.220 \\
1169.0316280 & V0486 Her                  & 12.8-13.6       & 12.70-13.25 & 89 & 0.161  & RRAB  & 0.8059    &  0.8061  &  0.8059  & 0.190 \\
             & ROTSE1 J172638.42+265616.5 & 13.551(0.497)R1 &             &    &        & RRAB  & 0.8056    &          &          &       \\
             & ASAS J172638+2656.3        & 12.769(1.202)   &             &    &        & CW-FU & 4.203     &          &          &       \\
1178.0358793 & ASAS J172907+2749.4        & 12.406(0.785)   & 12.43-13.31 & 93 & 0.274  & RRAB  & 0.46883   &  0.469   &  0.48871 & 0.310 \\ 
             & ROTSE1 J172907.35+274928.6 & 12.85-13.50(R1) &             &    &        & RRAB  & 0.46885   &          &          &       \\
1171.0322166 & 1RXS J172719.4+270858      & 12.9(0.12)      & 12.96-13.03 & 93 & 0.0143 & UV    & -         &  0.7486  &  0.7489  & 0.011 \\
1174.0340307 & V0420 Her                  & 14.5-15.6       & 14.37-15.18 & 86 & 0.179  & RRAB  & 0.6003    &  0.7464  &  0.5997  & 0.220 \\
1179.0338666 & [WM2007] 772               & 15.29(0.17)     & 14.99-15.27 & 93 & 0.053  & VAR   & -         &  0.1061  &  0.1107  & 0.026 \\
1167.0305285 & V0413 Her                  & 15.6-16.3       & 14.97-15.74 & 90 & 0.223  & RRC   & 0.5137    &  0.5131  &  0.5126  & 0.271 \\
1180.0314388 & V0879 Her                  & 15.2-15.8       & 15.24-16.06 & 82 & 0.175  & SXPhe & 0.0569    &  0.0569  &  0.0538  & 0.193 \\
1179.0338257 & [WM2007] 771               & 16.28(0.18)     & 15.94-16.36 & 67 & 0.091  & VAR   & -         &  0.9682  &  1.0172  & 0.04  \\
1169.0316908 & V0404 Her                  & 16.0-16.8       & 15.51-16.57 & 72 & 0.2276 & RR    & 0.55509   &  0.20918 &  0.6616  & 0.195 \\
1169.0316979 & V0405 Her                  & 15.8-17.0       & 15.76-16.79 & 47 & 0.2733 & RRAB  & 0.5879    &  0.5880  &  1.4287  & 0.296 \\    
\enddata
\label{known_var}
\end{deluxetable}

\subsection{Newly Found Variables} %3.3

Of the 67 previously uncatalogued variable candidates, 15 stars show
clearly perceived phased ightcurve patterns with periods well
determined by the LS algorithm.  Periods derived from the two LS
methods generally matched well. The remaining 52 candidates either did not
show any significant periods, perhaps because of insufficient phase
coverage of our data, or the phased lightcurves did not show patterns
readily recognizable. Table~\ref{classified_var} summaries the
properties of the 15 newly found variables. The first four columns
are the TAOS star identifier, RA and DEC coordinates taken
from the USNO-B1.0 catalog, both in
degrees, and the ID in the USNO catalog.  Columns 5 and 6 give the
calibrated TAOS magnitude and its error. The 7th and 8th columns are
the magnitude RMS of the lightcurve and the number of observations
used in the analysis. Columns 9 is the derived period and column 10 gives the
variable classification on the basis of the shape of lightcurves,
periods, and semi-amplitudes.

\begin{deluxetable}{lllllllcll}
\rotate
\tablecolumns{10} 
\tablecaption{Previously unknown variables}
\tabletypesize{\scriptsize}
\tablewidth{0pt}
\tablehead{ 
\colhead{ID} & \colhead{RA (J2000)} & \colhead{DEC (J2000)} &  \colhead{USNO ID} &  \colhead{TAOS\_mag} & \colhead{mag\_err} & \colhead{RMS} & \colhead{Frame No} & \colhead{Period} & \colhead{Classification}   \\
\colhead{} & \colhead{[Deg.]} & \colhead{[Deg.]} & \colhead{} &  \colhead{[mag]} & \colhead{[mag]} & \colhead{[mag]} & \colhead{} &\colhead{(days)} & \colhead{}} 
\startdata
TAOS 151-01 & 262.0523250 & +27.2035170 & 1172.0333288 & 11.330 & 0.005 & 0.0501 & 93 & 22.065 & SR1 \\
TAOS 151-02 & 261.7864306 & +27.0041889 & 1170.0320366 & 11.806 & 0.007 & 0.0251 & 93 & 59.891 & SR2 \\
TAOS 151-03 & 262.5683400 & +27.8393820 & 1178.0359464 & 12.503 & 0.009 & 0.0574 & 93 & 34.936 & SR3 \\
TAOS 151-04 & 261.6632700 & +27.3628750 & 1173.0332057 & 12.659 & 0.010 & 0.0356 & 92 & 2.409 & Cep  \\
TAOS 151-05 & 263.3984850 & +27.6555000 & 1176.0347892 & 12.869 & 0.011 & 0.0342 & 93 & 0.439 & RR1 \\
TAOS 151-06 & 262.7197200 & +27.6675860 & 1176.0346875 & 13.498 & 0.016 & 0.0850 & 92 & 0.147 & SXPhe \\
TAOS 151-07 & 262.3019550 & +27.6931360 & 1176.0346036 & 13.722 & 0.019 & 0.0823 & 93 & 0.280 & EW1 \\
TAOS 151-08 & 261.9320400 & +27.6780570 & 1176.0345244 & 14.088 & 0.023 & 0.0934 & 93 & 0.622 & RR2 \\
TAOS 151-09 & 263.3462100 & +26.7278840 & 1167.0306057 & 14.497 & 0.031 & 0.1040 & 93 & 0.401 & RR3 \\
TAOS 151-10 & 262.6746600 & +26.4792490 & 1164.0287321 & 14.619 & 0.033 & 0.1243 & 93 & 0.264 & Dsct \\
TAOS 151-11 & 263.1983550 & +27.1243050 & 1171.0324737 & 14.672 & 0.035 & 0.1985 & 93 & 0.174 & EW2 \\
TAOS 151-12 & 262.9039200 & +26.4763810 & 1164.0287687 & 15.121 & 0.048 & 0.1998 & 87 & 0.264 & EW3 \\
TAOS 151-13 & 262.8258600 & +26.4790000 & 1164.0287567 & 15.158 & 0.049 & 0.2028 & 93 & 0.264 & EW4 \\
TAOS 151-14 & 263.4367250 & +26.5488583 & 1165.0285251 & 15.235 & 0.052 & 0.1686 & 84 & 0.430 & EW5 \\
TAOS 151-15 & 263.2523100 & +27.1410000 & 1171.0324817 & 15.680 & 0.171 & 0.156 & 93 & 0.621 & RR4
\enddata
\label{classified_var}
\end{deluxetable}

The phased lightcurves of the 15 classified variables are shown in
Figure~5.  Among these, 4 are RR Lyrae (RR) variables, 3 are
semi-regular (SR) variables, 5 are broadly classified as eclipsing
binaries (EW), and one each has been classified as a $\delta$ Scuti
(Dsct), Cepheid (Cep) or SX Phonencis (SX Phe) variable.  

Cepheids (Cep) are massive stars with a spectral class of F at maximum
light while G to K at minimum; they generally have periods in the
range of 1--70~days with an amplitude variation of 0.1 to 2.0~mag in
V. RR Lyrae-type (RR) stars are radially pulsating giants with a
spectral class A--F; they have periods of 0.2 to 2.0~days with an
amplitude variation of 0.3 to 2.0~mag in V.  The $\delta$ Scuti-type
variable stars are A3-F0 main-sequence or sub-giant stars located in
the lower part of the classical instability strip in the H-R diagram,
with short pulsating periods ranging from 0.02 to 0.3 days and amplitudes less
than 0.1~mag in V.  The SX Phonencis (SX Phe) stars are pulsating
sub-dwarfs with a spectral type A2--F5.  Their light variations
resemble those of $\delta$ Scuti variables, but with shorter periods,
0.04 to 0.08 days, and larger magnitude variations, up to 0.7~mag in
V.  Semiregular (SR) variable are generally giants or supergiants of
intermediate or late (K-M) spectral types.  SRs show noticeable
periodicity in their lightcurves, with periods in the range from 20 to
1000 days and amplitudes varying in the range from 0.1 to 2.0~mag in
V.  Eclipsing binaries are binary systems with the orbital plane lying
near the line-of-sight of the observer.

Table~\ref{catalog_data} presents the USNO $B, R2$ and 2MASS $J, H,
K_s$ magnitudes of the previously known variable stars and the
variables found by our analysis, i.e., those listed in Table~\ref{known_var} and  
Table~\ref{classified_var}.  In addition to lightcurves, the color
information of the stars could be used to cross-check the
classification of variable stars.  TAOS data do not provide color
information, so we used the 2MASS data \citep{cut03} for this purpose.
The 2MASS observations were taken simultaneously in the $J, H$, and
$K_s$ bands, thus the colors $(J-H)$ and $(H-K_s)$ are sampled at the
same phase of a variable's light cycle.  The near-infrared colors of
RRab stars are in the range $(J-H)= -0.1$ to 0.5~mag and $(H-K_s) =
-0.1$ to 0.25~mag \citep{kin06}.  Figure~\ref{varjhk} displays the
2MASS colors of the variable stars listed in Table~\ref{catalog_data},
along with the loci of dwarfs, giants and supergiants \citep{bes98}
for reference.  With a few exceptions, the location of each variable
is reasonably positioned in the color-color diagram according to its
class.

\section{Summary and future work} %4

We have identified a total of 81 candidate variable stars  
in a particular field, No.\,151 (RA = 17$^{\rm h}$30$^{\rm m}$6$\fs67$,
Dec = 27\degr17\arcmin30\arcsec, J2000) in the TAOS survey.  
Among these, 29 variables can be classified, including 15 previously 
uncatalogued, as Cepheids, RR Lyrae stars, semiregular variables, 
eclipsing binaries or $\delta$ Scuti-type variables.
Their lightcurves, derived periods, semi-amplitudes, and hence the
variable classification are presented here and the data are avaiable 
on the TAOS website, {\verb+ http://taos.asiaa.sinica.edu.tw/demo+}.  
With the same methodology we expect 
to produce variable star lists in other TAOS fields, now with observations
covering more than 4 years (2005--2009).  In addition to stare-mode
photometry, the zipper-mode observations provide data sampled at 5~Hz,
so may be particularly useful for fast stellar variability \citep{kim10}.
The TAOS database hence has the unique potential to study several
thousand stars at timescales from less than a second to a few years.

\acknowledgments The work at National Central University was supported
by the grant NSC 96-2112-M-008-024-MY3.  YIB acknowledges the support
of National Research Foundation of Korea through Grant 2009-0075376.
Work at Academia Sinica was supported in part by the thematic research
program AS-88-TP-A02. Work at the Harvard College Observatory was
supported in part by the National Science Foundation under grant
AST-0501681 and by NASA under grant NNG04G113G. SLM's work was
performed under the auspices of the U.S. Department of Energy by
Lawrence Livermore National Laboratory in part under Contract
W-7405-Eng-48 and by Stanford Linear Accelerator Center under Contract
DE-AC02-76SF00515. KHC's work was performed under the auspices of the
U.S. Department of Energy by Lawrence Livermore National Laboratory in
part under Contract W-7405-Eng-48 and in part under Contract
DE-AC52-07NA27344.

\begin{deluxetable}{lllllll}
%\rotate
\tablecolumns{9} 
\tablecaption{Catalog for variable stars}
\tabletypesize{\scriptsize}
\tablewidth{0pt}
\tablehead{ 
 \colhead{USNO ID} & \colhead{$\Delta$diff} &  \colhead{USNOB2} & 
 \colhead{USNOR2} & \colhead{J}& 
 \colhead{ H}& \colhead{K$_s$}   \\
 \colhead{} &  \colhead{(")}& \colhead{[mag]}&\colhead{[mag]}&   \colhead{[mag]} &  \colhead{[mag]} &  \colhead{[mag]}} 
\startdata
& &   Known variables  & & & \\
\hline
1178.0360412 & 0.252 & 10.18 &  8.61  &  4.640 &  3.650 &  3.171        \\
1169.0319910 & 0.540 & 11.38 & 10.64  &  9.889 &  9.591 &  9.510        \\
1173.0334705 & 0.072 & 12.53 & 11.24  &  9.150 &  8.497 &  8.342        \\
1177.0363083 & 0.432 & 12.54 & 11.49  & 10.431 &  9.933 &  9.779        \\
1169.0316280 & 0.576 & 13.72 & 13.20  & 11.970 & 11.760 & 11.715        \\
1178.0358793 & 0.504 & 14.08 & 12.61  & 12.398 & 12.124 & 12.115        \\
1171.0322166 & 0.252 & 14.77 & 12.60  & 9.796  & 9.213  & 8.970         \\
1174.0340307 & 0.972 & 15.32 & 15.16  & 13.799 & 13.561 & 13.549        \\
1179.0338666 & 0.324 & 15.80 & 14.85  & 14.165 & 13.906 & 13.861        \\
1167.0305285 & 0.900 & ..... & 15.00  & 14.681 & 14.420 & 14.159        \\
1180.0314388 & 0.396 & 15.91 & 15.72  & 14.969 & 14.774 & 14.801        \\
1179.0338257 & 0.720 & 16.95 & 15.90  & 15.094 & 14.713 & 14.683        \\
1169.0316908 & 0.288 & 16.32 & 15.74  & 15.359 & 15.126 & 15.155        \\
1169.0316979 & 0.396 & 16.44 & 16.10  & 15.299 & 15.133 & 15.039        \\
\hline 
& &  Unknown variables &  & & \\
\hline
1172.0333288 & 0.864 & 12.64 & 11.02  &  8.264 &  7.392 &  7.129        \\
1170.0320366 & 0.828 & 14.47 & 11.24  & 10.491 & 9.901 &   9.852        \\
1178.0359464 & 0.540 & 14.61 & 12.64  &  9.097 &  8.221 &  7.947        \\
1173.0332057 & 0.180 & 13.63 & 12.83  & 11.360 & 10.852 & 10.715        \\
1176.0347892 & 0.756 & 13.18 & 12.11  & 11.830 & 11.486 & 11.416        \\
1176.0346875 & 0.648 & 14.12 & 13.25  & 12.211 & 11.731 & 11.632        \\
1176.0346036 & 0.468 & 14.69 & 13.91  & 12.796 & 12.479 & 12.406        \\
1176.0345244 & 0.612 & 15.19 & 14.12  & 13.134 & 12.902 & 12.835        \\
1167.0306057 & 1.440 & 15.01 & 14.59  & 14.071 & 13.906 & 13.840        \\
1164.0287321 & 0.252 & 15.96 & 14.92  & 13.404 & 12.843 & 12.747        \\
1171.0324737 & 0.252 & 15.43 & 14.86  & 13.428 & 13.063 & 13.039        \\
1164.0287687 & 0.468 & 17.31 & 15.46  & 13.453 & 12.824 & 12.594        \\
1164.0287567 & 0.180 & 16.56 & 15.64  & 14.701 & 14.319 & 14.343        \\
1165.0285251 & 7.056 & 20.09 & 18.65  & 16.687 & 16,198 & 15.663  \\
1171.0324817 & 0.684 & 16.31 & 15.88  & 14.738 & 14.372 & 14.403
\enddata
\label{catalog_data}
\end{deluxetable}

\begin{figure*}
 \plotone{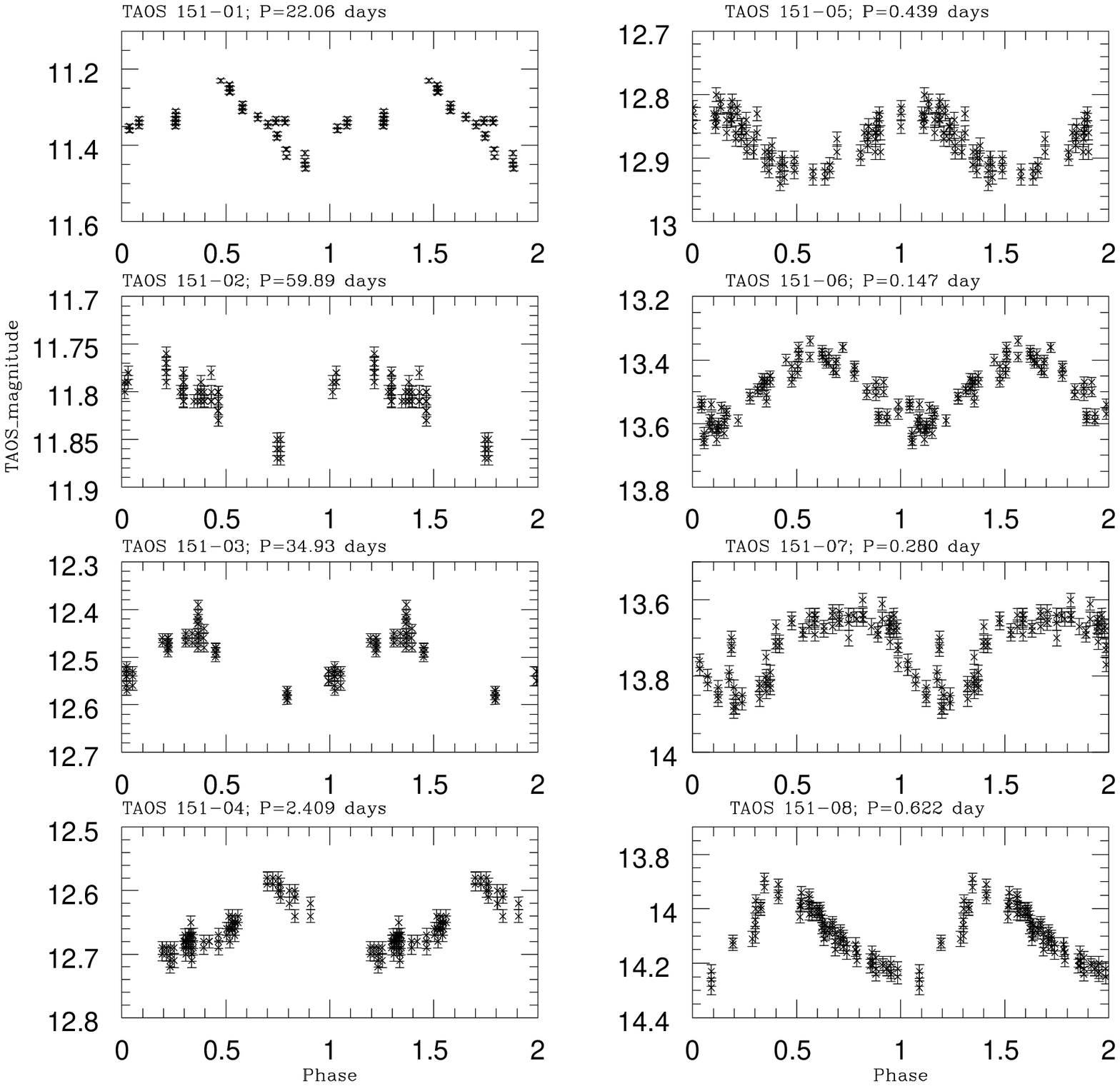}
\label{fig:phase}
\end{figure*}

%\addtocounter{figure}{-1}

\begin{figure}
 \plotone{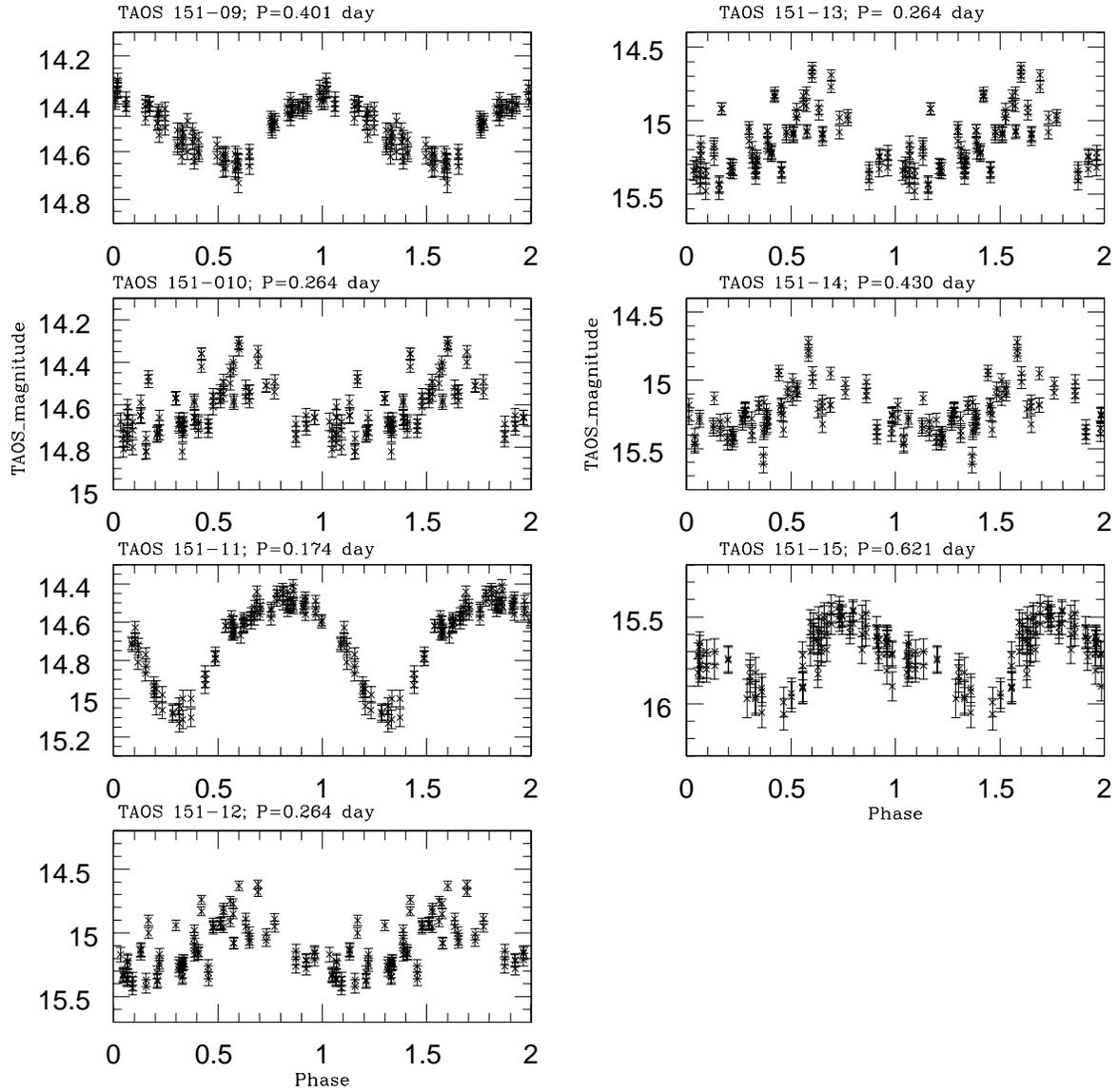}
  \caption{Phased lightcurves of newly identified variables with the TAOS data.}
\end{figure}

\begin{figure*}
%\plotone{figure6.ps} 
 \includegraphics[angle=-90, width=\textwidth]{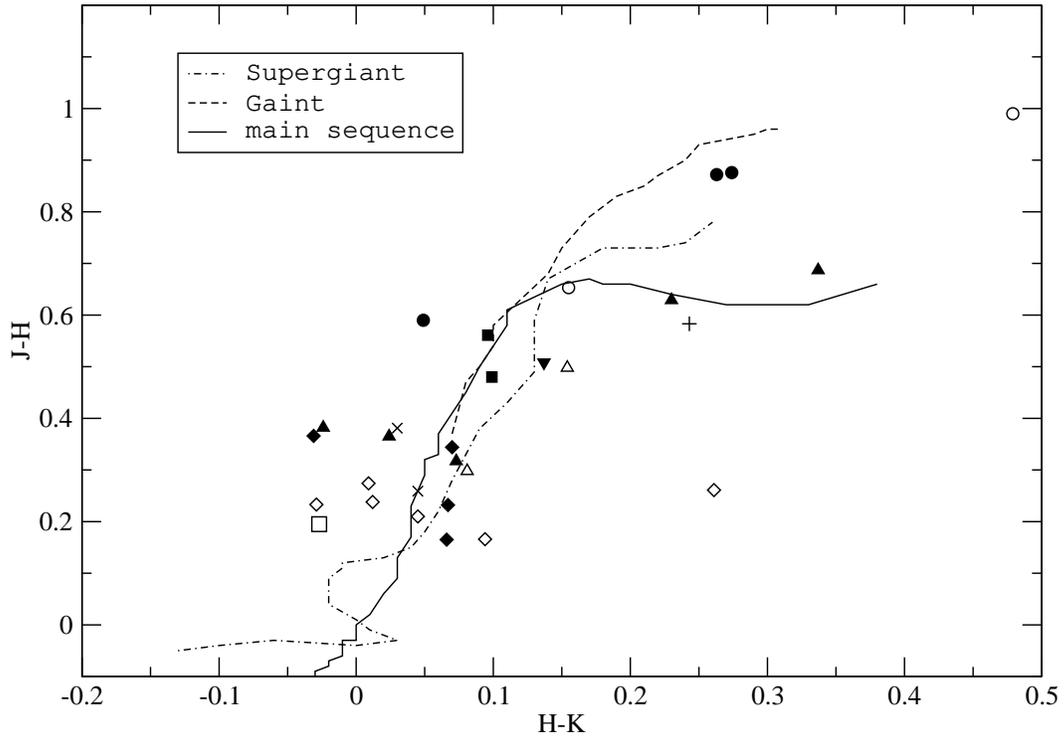}
 \caption{
 The 2MASS (J-H) versus (H-K$_s$) colors of variables stars.  The loci of dwarfs, giants and supergiants
 are taken from \citet{bes98}.  The open and filled symbols represent previously known and newly found
 variables.  Different symbols are for various variable classes: circles for SR, LPV, LB, diamonds 
 for RRAB, RRC, upward triangles for EW, EA, squares 
 for SX Phe, DSct, downward triangles for Cep, pluses for UV, and crosses for VAR.
         }
\label{varjhk}
\end{figure*}

\end{document}